\documentclass[english, journal=nalefd,manuscript=letter]{achemso}

\usepackage{graphicx}
\usepackage{color}
\usepackage[latin1]{inputenc}
\usepackage{amsbsy,amsmath,amssymb,amsthm,amsfonts}
\usepackage{times,mathptmx}
\usepackage{bm}
\usepackage{units}

\newcommand{\comment}[1]{}

\def \f{\textbf}

\def \<{\langle}
\def \>{\rangle}

\def \_{\underline}

\def \+{^{+}}
\def \-{^{\phantom{+}}}
\def \1{\phantom{,}^1}
\def \2{\phantom{,}^2}
\def \3{\phantom{,}^3}

\title{Graphene as Gain Medium for Broadband Lasers}

\author{Roland Jago, Torben Winzer, Andreas Knorr, and Ermin Malic}
\email{ermin.malic@tu-berlin.de}
\affiliation{
University of Technology Berlin, Department for Theoretical Physics, Non-linear Optics and Quantum Electronics, Hardenbergstasse 36, 10623 Berlin, Germany}

\begin{document}

\begin{abstract}

In contrast to conventional structures, efficient non-radiative carrier recombination counteracts the appearance of optical gain in graphene. Based on a microscopic and fully quantum-mechanical study of the coupled carrier, phonon, and photon dynamics in graphene, we present a strategy to obtain a long-lived gain: Integrating graphene into a photonic crystal nanocavity and applying a high-dielectric substrate gives rise to pronounced coherent light emission suggesting the design of graphene-based laser devices covering a broad spectral range.

\end{abstract}

Visible and infrared fiber lasers are the basis for information technology, while microwave and radio-frequency emitters build the backbone of wireless communications. Due to the lack of efficient sources of terahertz light, there is a technological gap between these two frequency ranges  \cite{koehler02}. The ongoing search for novel gain materials has brought ultrathin layered nanomaterials into the focus of current research. Graphene, a single layer of carbon atoms, exhibits a linear and gapless band structure around the Dirac point \cite{novoselov04, geim07}. This unique electronic dispersion offers a broad spectrum of optically active states including the terahertz region. This extraordinary feature has already been technologically exploited in graphene-based photodetectors covering a wide range of frequencies \cite{xia09, mueller10, echtermeyer11, furchi12} as well as graphene-based saturable absorbers converting the continuous wave output of lasers into a train of ultrashort optical pulses \cite{
ferrari10, bonaccorso10, avouris12}. Recently, hot photoluminescence has been observed across the entire visible spectrum even exceeding the energy of the excitation reflecting radiative recombination of excited carriers \cite{lui10, liu10, stoehr10, khanh13}. In the strong excitation regime, a spectrally broad population inversion has been measured \cite{li12,gierz13} and theoretically predicted \cite{ryzhii07,winzer13}. The excited carriers become quickly redistributed filling the optically active states in the vicinity of the Dirac point, where the reduced density of states gives rise to a relaxation bottleneck and a build-up of a population inversion resulting in optical gain. However, non-radiative carrier recombination channels reduce the accumulation of charge carriers accounting for a decay of the population inversion on a femtosecond time scale \cite{winzer13}  making it unsuitable for technological applications. 

\begin{figure}[!t]
\begin{centering}
\includegraphics[width=95mm]{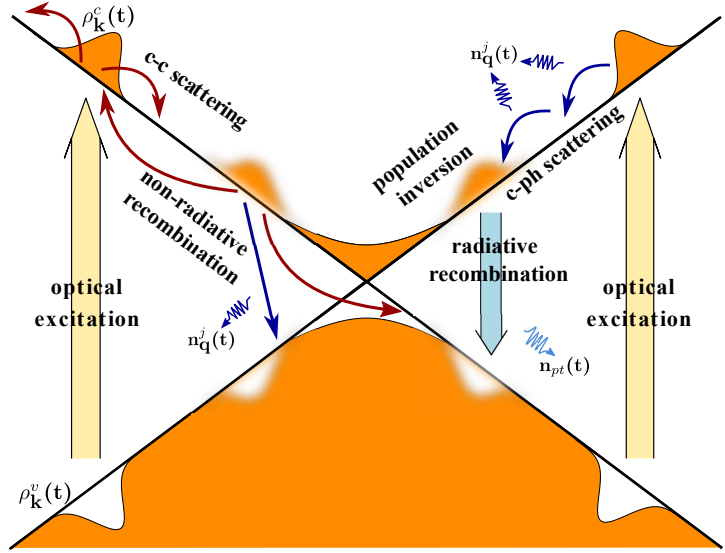} 
\par\end{centering}
\caption{\textbf{Schematic illustration of the carrier, phonon, and photon dynamics in optically pumped graphene.}  Non-equilibrium carriers relax towards lower energies via carrier-carrier (c-c) and carrier-phonon (c-ph) scattering. Above a threshold pump strength, a population inversion occurs giving rise to an increased efficiency for radiative recombination. However, at the same time, the accumulation of carriers also enhances the competing non-radiative recombination channels reducing the achieved population inversion.
 }
\label{fig1_sketch} 
\end{figure}

In this Article, we address the question whether a long-lived optical gain can be achieved in graphene. This presents the key prerequisite for the realization of graphene-based broadband laser devices that could also operate in the technologically relevant terahertz spectral region. To answer this question, we perform for the first time a microscopic and fully quantum-mechanical investigation of the coupled carrier, phonon, and photon dynamics in strongly pumped graphene within a cavity.  
We shed light on the occurring carrier-carrier, carrier-phonon, and carrier-photon interactions on the same microscopic footing allowing us to track the way of non-equilibrium carriers resolved in time and energy. The performed microscopic treatment of Coulomb-induced and phonon-assisted relaxation processes as well as radiative and non-radiative carrier recombination channels unravels the driving microscopic mechanism underlying the emission of light from graphene:  Fig. \ref{fig1_sketch} shows a sketch of the competing relaxation processes for optically pumped carriers. The goal is to predict a regime characterized by weak non-radiative recombination processes and strong carrier-light interaction resulting in a long-lived gain and pronounced light amplification. 
Here, we propose to enhance the carrier-light coupling by integrating graphene into a planar photonic crystal nanocavity \cite{akahane03, furchi12, engel12, gan12} and at the same time to reduce the efficiency of the predominant Coulomb-induced non-radiative recombination by considering graphene on a substrate inducing a large dielectric screening. To address the quantum statistics of the emitted light and to answer the question whether coherent laser light can be achieved, we determine the temporal evolution of the second-order autocorrelation function.\\

\textbf{Theoretical approach:} 
The study is performed within the formalism of the  density matrix theory \cite{kira99, haug04, rossi02, gies07} offering microscopic access to temporally and spectrally resolved carrier, phonon, and photon dynamics \cite{erminbuch, malic11b}. The many-particle Hamilton operator $H=H_0+H_{\text{c,c}}+H_{\text{c,p}}+H_{\text{c,pt}}$ consists of the  (i)  free-carrier,  phonon, and photon contribution $H_0$,  (ii)  carrier-carrier $H_{\text{c,c}}$ and (iii) carrier-phonon interaction $H_{\text{c,p}}$ accounting for Coulomb- and phonon-induced intraband scattering and non-radiative recombination processes, respectively,  and finally (iv) carrier-photon interaction $H_{\text{c,pt}}$ describing the radiative recombination of excited carriers, cf. Fig. \ref{fig1_sketch}.

Treating all many-particle interactions on the same quantum-mechanical level, we derive graphene luminescence equations - a coupled set of differential equations for the carrier occupation probability $\rho_{\f{k}}^\lambda (t)=\langle a_{\f{k}\lambda}^{+}a^{\phantom{+}}_{\f{k}\lambda}\rangle$ in the valence ($\lambda = v$) and the conduction band ($\lambda = c$) with the electronic momentum $\bf k$, the phonon number $n_{\f{q}}^j(t)=\langle b_{\f{q} j}^{+}b^{\phantom{+}}_{\f{q} j}\rangle$ for different optical and acoustic phonon modes $j$ with the phonon momentum $\bf q$, and finally the photon number $n_{pt}^\gamma(t)=\langle c_{\gamma}^{+}c^{\phantom{+}}_{\gamma}\rangle$  and the photon-photon correlation $n_{pt,2}^\gamma(t)=\langle c_{\gamma}^{+} c_{\gamma}^{+} c^{\phantom{+}}_{\gamma} c^{\phantom{+}}_{\gamma}\rangle^{c}$ in the mode $\gamma$ determined by the cavity. Here, we have introduced $a_{\f{k}\lambda}^+$ and
$a^{\phantom{+}}_{\f{k}\lambda}$ as creation and annihilation operators for carriers,  $b_{\f{q}j}^+$ and
$b^{\phantom{+}}_{\f{q}j}$ for phonons, and $c_{\gamma}^+$ and
$c^{\phantom{+}}_{\gamma}$ for photons, respectively. More details are presented in the supplementary material and Refs. \cite{erminbuch, malic11b}.

Here, we focus on the quantum optical aspect of the theory and discuss in more detail the Hamilton operator $H_{\text{c,pt}}$ describing the carrier-photon interaction:
\begin{equation}
\label{h_cpt}
H_{\text{c,pt}}=i\hbar\sum_{\f{k},\gamma}\tilde{M}^{vc}_{\gamma\f{k}}(a^+_{\f{k}v}a^{\phantom{+}}_{\f{k}c}c^{+}_\gamma - a^+_{\f{k}c}a^{\phantom{+}}_{\f{k}v}c^{\phantom{+}}_\gamma) \quad\quad \text{with}\quad\quad  
 \tilde{M}_{\gamma\bf k}^{vc}= \frac{e_0}{m_0}\sqrt{\frac{\hbar}{2\omega_{0}\varepsilon_{0}V}} \,\hat{\bf{e}}_\gamma\cdot\bf{M}^{vc}_{\f{k}}\,.
 \end{equation}
The carrier-photon matrix element $\tilde{M}_{\gamma\bf k}^{vc}
$ determines the strength of the interaction. It depends on the properties of the optical cavity characterized by a dominant mode $\gamma$ with the eigen frequency $\omega_0$ and the cavity volume $V$. In particular, the carrier-photon interaction is the strongest for small cavity frequencies and small cavity volumina. 
Furthermore, the coupling has the same symmetry as the semi-classical optical matrix element ${\bf{M}}^{vc}_{{\bf{k}}}=  \langle \f{k}v | \nabla| \f{k}c \rangle$ \cite{malic11b} and depends on the polarization direction of the excitation pulse $\hat{\bf e}_\gamma$.
The emitted light intensity from optically pumped graphene is characterized by the photon dynamics $\dot{n}_{pt}^\gamma(t)$ driven by the photon-assisted transition $\1S^{vc}_{\gamma\bf k}=\langle a_{\f{k}v}^{+}a^{\phantom{+}}_{\f{k}c} c_{\gamma}^+  \rangle^c$. The latter describes the recombination of excited carriers between the conduction and the valence band in the state $\bf k$ accompanied by the emission of a photon in the mode $\gamma$, cf. the supplementary material. We investigate the influence of the photon dynamics on the temporal evolution of the carrier and phonon occupations $\rho_{\bf k}^\lambda (t),\, n_{\bf q}^j(t)$ and vice versa. In particular, we study the interplay between the many-particle processes of radiative and non-radiative carrier recombination, which turns out to be crucial for the realization of graphene-based laser devices. 

To close the hierarchy of many-particle equations, the Coulomb-induced and phonon-assisted many-particle interactions are treated within the second-order Born-Markov approximation \cite{haug04, erminbuch}. The carrier-photon coupling is considered up to the forth-order in the Born approximation to have access to the photon statistics described by the second-order autocorrelation function \cite{scully97}: 
\begin{equation}
 g^{(2)}_\gamma (t)=\frac{\langle c_{\gamma}^+ c_{\gamma}^+ c^{\phantom{+}}_{\gamma} c^{\phantom{+}}_{\gamma} \rangle (t)}{\langle c_{\gamma}^+ c^{\phantom{+}}_{\gamma} \rangle^2(t)}=2+\frac{\langle c_{\gamma}^+ c_{\gamma}^+ c^{\phantom{+}}_{\gamma} c^{\phantom{+}}_{\gamma} \rangle^c (t)}{\langle c_{\gamma}^+ c^{\phantom{+}}_{\gamma} \rangle^2(t)}=2+\frac{n^\gamma_{pt,2}(t)}{(n^\gamma_{pt}(t))^2}.
\label{g2}
\end{equation}
Taking into account the full dynamics of the appearing photon-photon correlations $n^\gamma_{pt,2}(t)$, we obtain a closed set of coupled graphene luminescence equations determining the photon dynamics and the quantum statistics of light emitted from graphene within a cavity, cf. the supplementary material. In particular, having access to the second-order autocorrelation function, we are able to distinguish coherent laser light characterized by the Poisson statistics with  $g_{\gamma}^{(2)}=1$ from thermal light  ($g_{\gamma}^{(2)}>1$) and non-classical light ($g_{\gamma}^{(2)}<1$) \cite{scully97}. More details on the derived equations of motion can be found in the supplementary material.\\
\begin{figure}[!t]
\begin{centering}
\includegraphics[width=140mm]{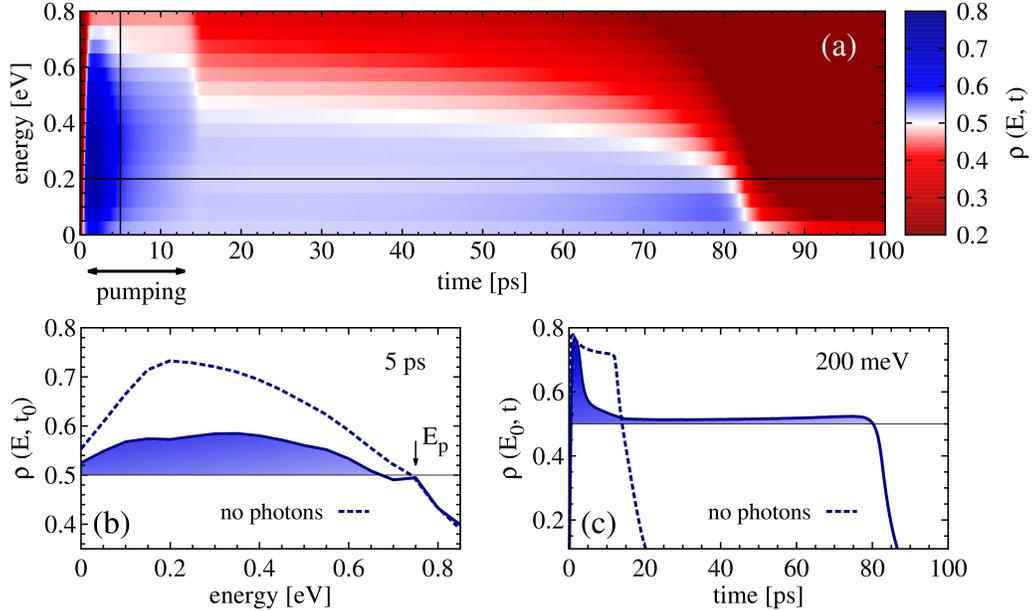}
 \par\end{centering}
\caption{\textbf{Temporally and spectrally resolved carrier dynamics.}  (a) The surface plot shows the non-equilibrium dynamics of optically pumped carriers in the conduction band. Blue area displays the region of population inversion with $\rho(E,t)>0.5$. The carrier occupation is shown  (b) as a function of the carrier energy  for the fixed time of \unit[5]{ps} after the switch-on of the optical pumping pulse and (c) as a function of time at the fixed carrier energy of \unit[200]{meV} corresponding to the investigated cavity energy. 
The width of the pump pulse is \unit[10]{ps} and the pumping energy $E_{p}$ is \unit[1.5]{eV} corresponding to a carrier energy of \unit[0.75]{eV}.  The dashed lines show the carrier dynamics without photons illustrating the crucial impact of the carrier-photon coupling on the dynamics of carriers.}
\label{fig2_electrons} 
\end{figure}

\textbf{Coupled carrier, phonon, and photon dynamics:} 
After solving the graphene luminescence equations, we have full microscopic access to time- and energy resolved carrier, phonon, and photon dynamics in graphene within a cavity. A non-equilibrium carrier distribution is achieved by  pumping graphene with a symmetric steady-state-like pulse 
characterized by the pumping rate of $P_0=\unit[1]{fs^{-1}}$, the pump energy of $E_0=\unit[1.5]{eV}$, and spectral and temporal widths of \unit[0.02]{eV} and \unit[10]{ps}, respectively. We assume the graphene layer to be integrated in a planar photonic crystal nanocavity, as already successfully realized by X. Gan and co-workers \cite{gan12}. Such nanocavities are known to have the potential to reach extremely high quality factors $Q$ and ultrasmall cavity volumes $V$ \cite{akahane03}. For this study, we assume $Q=45000$ and $V=\unit[7\cdot10^{-14}]{cm^3}$ according to the work of Y. Akahane and co-workers \cite{akahane03}. Note that the main message of this work remains valid even for smaller quality factors and larger cavity volumes, as discussed below. Furthermore, we investigate the emission of photons in the cavity mode of \unit[400]{meV} corresponding to the carrier energy, where the maximal population inversion is reached. Later on, we will decrease the cavity energy down to few meV to access the 
terahertz frequency range. Finally, to partially suppress non-radiative recombination processes we assume that graphene lies on a substrate with a large dielectric constant $\varepsilon$. Here, we have exemplary considered hafnium dioxide with a dielectric constant of $\varepsilon=25$ \cite{wilk01}. We take into account that graphene is surrounded by air on the one side and by the substrate on the other by averaging the background dielectric constant with $\varepsilon=(\varepsilon_{\text{air}}+\varepsilon_{\text{substrate}})/2
$ \cite{louie09}.

First, we discuss the time- and energy-resolved dynamics of non-equilibrium carriers and phonons after optical pumping. Having established a long-lived population inversion, we then investigate the dynamics of photons and of photon-photon correlations determining the quantum statistics of emitted light. Figure \ref{fig2_electrons} shows the temporal and spectral evolution of optically pumped carriers in graphene within a photonic crystal nanocavity. We observe a pronounced spectrally broad population inversion, i.e. the occupation of carriers in the conduction band $\rho(E,t)$ is higher than 0.5 (blue region in Fig. \ref{fig2_electrons}(a)). Our calculations reveal that the non-equilibrium carriers are quickly scattered to energetically lower states via carrier-carrier and carrier-phonon interactions leading to an ultrafast accumulation of carriers. Already a pump pulse with a width of \unit[10]{ps} provides enough electrons to achieve a  pronounced and spectrally broad population inversion up to energies 
slightly below the pump energy $E_p$, cf. Fig. \ref{fig2_electrons}(b). 

To study the dynamics of the microscopic mechanism behind the population inversion in more detail, we show in Fig. \ref{fig2_electrons}(c) the carrier occupation as a function of time at the fixed carrier energy of \unit[200]{meV} (corresponding to the investigated cavity energy of \unit[400]{meV}) with and without the carrier-photon coupling.
In the full calculation including the impact of photons (solid line), the population inversion is long-lived, i.e. in the range of \unit[100]{ps}.
We find pronounced carrier occupation values of up to 0.8 as long as the pump pulse reaches its maximum. Then, the occupation quickly decreases predominantly due to the efficient radiative recombination. The loss of carriers via radiative and non-radiative recombination processes can be partially compensated during and after the pumping by intraband scattering from energetically higher states. The interplay of radiative emission and absorption processes on the one side and non-radiative recombination channels and Coulomb- and phonon-induced intraband scattering processes on the other side results in a quasi-equilibrium keeping the carrier occupation at a stationary value slightly above 0.5 on a time scale of \unit[100]{ps}. Once the loss of carriers cannot be compensated anymore, the carrier occupation quickly drops to small values determined by the thermal distribution. Applying stronger and longer excitation pulses, we can further increase the life time of the population inversion.

The crucial role of radiative processes is demonstrated in calculations neglecting the dynamics of photons (dashed line in Fig. \ref{fig2_electrons} (c)). Here, we find a decay of the population inversion on a time scale of  the pump pulse. In the absence of light emission, non-radiative recombination channels are predominant giving rise to an ultrafast reduction of the population inversion. This behavior is well reflected by the corresponding radiative and non-radiative recombination rates, cf. the Figure 1 in the the supplementary material. 
A necessary condition for a long-lived population inversion is a considerable suppression of the efficient processes of Auger recombination \cite{winzer10, wendler14}. These Coulomb-induced interband processes bringing excited electrons down to the valence band  are known to be by orders of magnitude higher than radiative recombination in free-suspended graphene \cite{winzer13}. Therefore, in this work, we introduce a substrate with a large dielectric constant, which considerably reduces the efficiency of Coulomb-induced non-radiative processes. Furthermore, implementing graphene into a photonic crystal nanocavity, we strongly enhance the carrier-light interaction. With this recipe, we achieve radiative recombination rates that are two orders of magnitude higher than the competing non-radiative processes, cf. Figure 1 in the supplementary material.

\begin{figure}[!t]
\begin{centering}
\includegraphics[width=150mm]{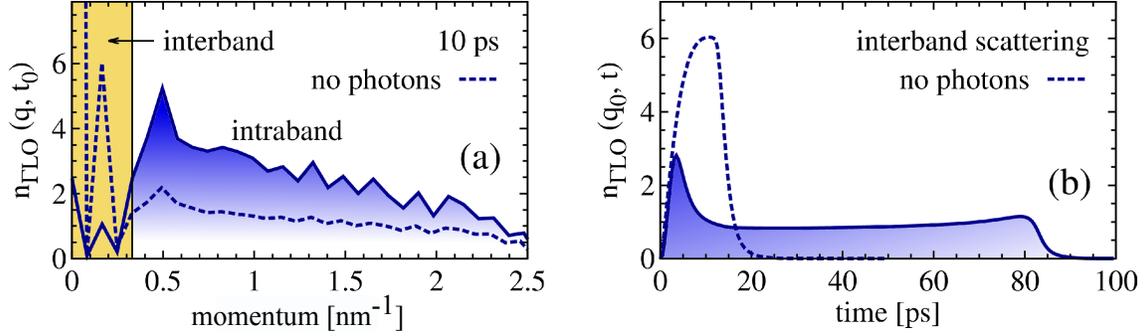}
 \par\end{centering}
\caption{\textbf{Temporally and spectrally resolved phonon dynamics.} (a) Spectral resolution of the exemplary $\varGamma LO$ phonons at a fixed time of \unit[10]{ps} after the switch-on of the optical excitation. The orange-shaded region emphasizes phonons with the momentum $q<\unit[0.3]{nm^{-1}}$ inducing interband scattering of excited carriers (non-radiative recombination), whereas phonons with $q>\unit[0.3]{nm^{-1}}$ contribute to intraband scattering \cite{malic11b}. The dashed lines show the dynamics without photons revealing more pronounced recombination on the one hand and considerably suppressed intraband processes on the other hand. (b) Temporal resolution of the $\varGamma LO$ phonon dynamics at a fixed momentum of $\unit[0.18]{nm^{-1}}$ describing the efficiency of non-radiative carrier recombination induced by the emission of $\varGamma LO$ phonons. }
\label{fig4_phonons} 
\end{figure}

The dynamics of optically pumped carriers is significantly influenced by efficient intra- and interband carrier-phonon scattering. Our calculations contain the contribution of all relevant optical and acoustic phonon modes ($\varGamma LO, \varGamma TO, KTO, \varGamma LA$).  Figure
 \ref{fig4_phonons} reveals the temporally and spectrally resolved dynamics of the exemplary $\varGamma LO$ phonons. Compared to the initial Bose-Einstein distribution at room temperature, we find a considerable non-equilibrium phonon distribution over a broad momentum range reflecting significant intraband carrier scattering via emission of hot $\varGamma LO$ phonons. The orange-shaded region in Fig. \ref{fig4_phonons}(a) illustrates the phonon-assisted interband scattering resulting in a non-radiative recombination of carriers. The temporal evolution of the corresponding $\varGamma LO$ phonons shown in Fig. \ref{fig4_phonons}(b) reproduces well the behavior found for the phonon-assisted carrier recombination rate (cf. Figure 1(b) in the supplementary material): The phonon occupation reaches a quasi-stationary state on a time scale of \unit[100]{ps}.  This reflects the compensation of the radiatively induced carrier loss through the supply of further carriers via intraband scattering from higher electronic 
states. Considering the phonon dynamics without the impact of photons, we find that the interband scattering is significantly enhanced, while at the same time the efficiency of intraband processes is reduced, cf. the dashed line in Fig. \ref{fig4_phonons}(a). The emission of light is in direct competition to phonon-assisted interband scattering. It depopulates the states close to the Dirac point, which in turn drives the phonon-assisted intraband scattering to fill these states. This explains the weaker intraband processes in the absence of the carrier-photon interaction.\\
 
\textbf{Photon dynamics and quantum statistics of emitted light:}
Having understood the carrier and phonon dynamics, we now focus on the temporal evolution of emitted photons from optically pumped graphene within a nanocavity. We investigate the photon dynamics $n_{pt}(t)$ depending on the width of the applied pump pulse and depending on the cavity energy including the technologically relevant terahertz spectral range. 
\begin{figure}[!t]
\includegraphics[width=85mm]{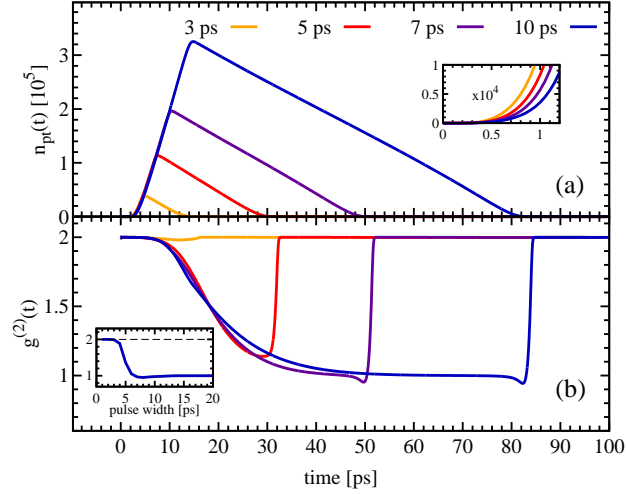} 
\caption{\textbf{Photon dynamics and quantum statistics.} Illustration of the temporal evolution  (a) of photons emitted from optically pumped graphene within a photonic crystal nanocavity and (b) of the second-order autocorrelation function $g^{(2)}(t)$ expressing the quantum statistics of emitted light for different widths of the applied pump pulse (at a constant pump rate $P_0$).   The cavity energy is fixed to \unit[400]{meV}.
The inset in (a) corresponds to a zoom-in revealing the turn-on dynamics. 
The inset in (b) shows the minimum of $g^{(2)}$ as a function of the pulse width revealing that coherent laser light characterized by  $g^{(2)}(t)=1$ occurs for excitations longer than approximately \unit[7]{ps}.}
\label{fig5_photons} 
\end{figure}
Figure \ref{fig5_photons}(a) shows $n_{pt}(t)$ at the fixed cavity energy of \unit[400]{meV} for different pump pulse widths modeling the behavior towards stationary pumping.   In the first few picoseconds, the number of emitted photons increases very slowly and can be ascribed to the spontaneous emission, cf. also the inset of Fig. \ref{fig5_photons}(a). After this characteristic delay time determined by the pump strength, the process of induced emission boosts the number of photons as long as the pump pulse is switched on. The pumping maintains the carrier supply that is necessary to keep the population inversion on a high level (Fig. \ref{fig2_electrons}(c)). After pumping, the number of photons gradually decreases directly reflecting the behavior of the radiative recombination rate, cf. Figure 1(a) in the supplementary material.  As soon as the carrier loss cannot be compensated anymore by Coulomb- and phonon-induced in-scattering of further carriers from energetically higher states, the number of 
photons quickly drops, since the population inversion vanishes and the stimulated emission does not contribute anymore. Figure \ref{fig5_photons}(a) reveals that the decay of the number of emitted photons can be controlled by the length of the optical excitation. The longer the pump pulse, the more stable is the population inversion resulting in a more pronounced emission of light.

To investigate whether the emitted light corresponds to coherent laser light, we calculate the second-order autocorrelation function $g^{(2)}$  allowing us to determine the photon statics. Perfectly coherent laser light follows the Poissonian photon statistics characterized by $g^{(2)}=1$ describing random time intervals between the emitted photons \cite{scully97}. Figure \ref{fig5_photons}(b) shows the temporal evolution of $g^{(2)}$ for different pump widths. As long as the photon dynamics is dominated by the processes of spontaneous emission, we find $g^{(2)}\approx 2$ corresponding to the value expected for thermal chaotic light. 
After few picoseconds, $g^{(2)}$ starts to decrease reaching the value of 1 after few tens of ps, i.e. optically excited graphene within a nanocavity indeed emits coherent laser light. Whether perfectly coherent light can be obtained, depends on the duration of the pumping. For pump pulse widths larger than approximately \unit[7]{ps}, $g^{(2)}=1$ can be reached.

\begin{figure}[!t]
\includegraphics[width=100mm]{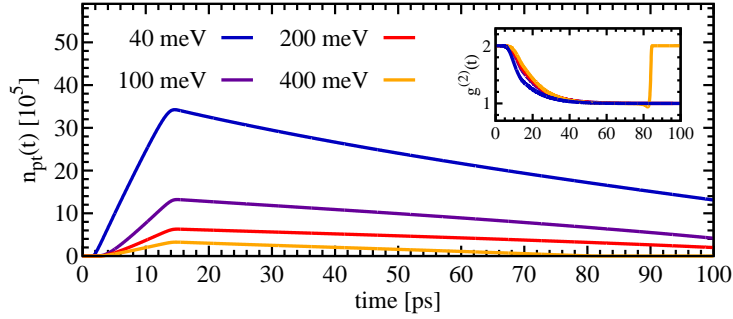} 
\caption{\textbf{Photon dynamics for different cavity energies.} Temporal evolution of emitted photons for different cavity energies covering a broad spectral range at a fixed pump pulse width of \unit[10]{ps}. The inset shows the corresponding  second-order autocorrelation function $g^{(2)}(t)$. The most pronounced dynamics is found at the smallest energy suggesting efficient emission of coherent terahertz light. 
}
\label{fig6_cavity_energy} 
\end{figure}

So far, we have investigated the emission of photons in the infrared spectral range. Figure \ref{fig6_cavity_energy} illustrates the dynamics of photons with decreasing cavity energy realized by changing the geometry of the cavity at a fixed pump width and quality factor of the cavity.  The qualitative behavior of the emitted photons and the corresponding second-order autocorrelation function (shown in the inset of Fig. \ref{fig6_cavity_energy}) is similar, however, the quantitatively most pronounced and the longest lasting emission of coherent light is achieved at the smallest investigated energy of \unit[40]{meV}. We find an increase of the photon number by a factor of 10 compared to the cavity energy of \unit[400]{meV}. Consequently, $n_{pt}(t)$ drops to zero at later times resulting in emission of coherent laser light (characterized by $g^{(2)}=1$) for much more than \unit[100]{ps}. The reason lies in the increased carrier-photon coupling due to the inverse dependence of the coupling element $\tilde{M}_{\gamma\bf k}^{vc}$ on the cavity energy $\hbar\omega_0$, cf. Eq. (1). 
Going further to the range of just few meV, we expect even a much more pronounced emission of coherent laser light suggesting the design of graphene-based terahertz laser devices.

In summary, based on a microscopic and fully quantum-mechanic study we have gained new fundamental insights into the coupled carrier, phonon, and photon dynamics in optically pumped graphene within a photonic crystal nanocavity. We have revealed that under certain experimentally controllable conditions a long-lived optical gain can be achieved suggesting graphene as a novel material for laser devices operating at a broad range of frequencies.  The obtained microscopic insights shed light on elementary many-particle processes behind the amplification of light in graphene presenting a crucial step towards the realization of future graphene-based nanolasers.

We acknowledge financial support from the German Science Foundation through the priority program SPP 1459 (E.M.), the collaborative research center SFB 787 (A.K.), and the Research Training Group GRK 1558 (R. J.). 
Furthermore, E.M. thanks the Einstein Foundation Berlin. 

%\bibliography{papers}

\providecommand*{\mcitethebibliography}{\thebibliography}
\csname @ifundefined\endcsname{endmcitethebibliography}
{\let\endmcitethebibliography\endthebibliography}{}

\end{document}